\documentstyle[emulate_apj,apjfonts,epsf]{article}
\gdef\h50min{$h_{50}^{-1}$}
\gdef\1054{MS\,1054$-$03}
\gdef\tstart{t_{\rm start}}
\gdef\fstar{f_*}
\gdef\tstop{t_{\rm stop}}
\gdef\taustop{\tau_{\rm stop}}
\gdef\t0{t_0}
\lefthead{van Dokkum \& Franx}
\righthead{Early-Type Galaxies in Clusters}
\slugcomment{Accepted for publication in The Astrophysical Journal}
\begin{document}
\title{Morphological Evolution and the Ages of Early-Type
Galaxies in Clusters}
\author{Pieter G. van Dokkum\altaffilmark{1}}
\affil{California Institute of Technology, MS105-24, Pasadena,
CA 91125}
\and
\author{Marijn Franx}
\affil{Leiden Observatory, P.O. Box 9513, NL-2300 RA, Leiden, The
Netherlands}
\altaffiltext{1}{Hubble Fellow}

\begin{abstract}

Morphological and spectroscopic
studies of high redshift clusters indicate that a
significant fraction of present-day early-type galaxies was transformed
from star forming galaxies at $z<1$. On the other hand, the slow
luminosity evolution of early-type galaxies and the low scatter in
their color-magnitude relation indicate a high formation redshift of
their stars.  In this paper we construct models
which reconcile these apparently
contradictory lines of evidence, and we quantify the effects of morphological
evolution on the observed photometric properties of early-type
galaxies in distant clusters. We show that in the case of strong
morphological evolution the apparent luminosity and color evolution of
early-type galaxies are similar to that of a single age stellar
population formed at $z= \infty$, irrespective of the true star
formation history of the galaxies.  Furthermore, the scatter in age,
and hence the scatter in color and luminosity, is approximately
constant with redshift.  These results are consequences of the
``progenitor bias'': the progenitors of the youngest low redshift
early-type galaxies drop out of the sample at high redshift. 
We construct models which
reproduce the observed evolution of the number fraction of early-type
galaxies in rich clusters and their color and luminosity evolution
simultaneously.  Our modelling indicates that $\sim 50$\,\% of
early-type galaxies were transformed from other
galaxy types at $z<1$, and their progenitor
galaxies may have had roughly constant star formation rates prior to
morphological transformation.
The effect of the progenitor bias on the evolution of
the mean $M/L$ ratio and color can be estimated.
The progenitor bias
is a linear function of the scatter in the color-magnitude relation
produced by age variations, and is maximal if the observed scatter
is entirely due to age differences. After correcting the
observed evolution of the mean $M/L_B$ ratio for the maximum
progenitor bias we find that the mean
luminosity weighted formation redshift of stars in early-type galaxies
$\langle z_* \rangle = 3.0^{+0.9}_{-0.5}$ for $\Omega_m=0.3$ and
$\Omega_{\Lambda}=0$, and $\langle z_* \rangle = 2.0^{+0.3}_{-0.2}$ for
$\Omega_m=0.3$ and $\Omega_{\Lambda}=0.7$.
Our analysis places the star formation epoch of early-type galaxies
later than previous studies which ignored the effects of progenitor bias.
The results are consistent with the idea that
(some) Ly-break galaxies are star forming building blocks of
massive early-type galaxies in clusters.

\end{abstract}

keywords{
galaxies: evolution,
galaxies: elliptical and lenticular, cD, galaxies: structure of,
galaxies: clusters
}

\section{Introduction}

Early-type galaxies (elliptical and S0 galaxies)
constitute $\sim 80$\,\% of the galaxy population
in the central regions of nearby rich clusters (Dressler 1980), and
studies of nearby and distant clusters have provided
strong constraints on their evolution and formation.

The cluster observations have been used to constrain the stellar
ages of early-type galaxies.
On one hand, there is very good evidence that
early-type galaxies form a very
homogeneous, slowly evolving population: the scatter in their colors is
small, both at low redshift (e.g., Bower, Lucey, \& Ellis 1992), and
at high redshift (Ellis et al.\ 1997; Stanford, Eisenhardt, \&
Dickinson 1998), the scatter in their mass-to-light ratios is equally
small
(e.g., Lucey et al.\ 1991, Pahre, Djorgovski, \& de Carvalho 1998,
Kelson et al.\ 2000), and the evolution of
their mass-to-light ratios is slow (van Dokkum \& Franx 1996, Kelson et
al.\ 1997, Bender et al.\ 1998, van Dokkum et al.\ 1998a). These
results indicate a very high redshift of formation for the stars in
early-type galaxies. As an example, van Dokkum et al (1998a) found
from the evolution of the Fundamental Plane relation to $z=0.83$ that
the stars in massive early-type galaxies must have formed at $z>2.8$
for $\Omega_m =0.3$, $\Omega_{\Lambda}=0$, and a Salpeter (1955) IMF.
Furthermore the low scatter in colors and $M/L$ ratios
implies that the scatter in ages at any redshift is very small (e.g.,
Ellis et al.\ 1997, Stanford et al.\ 1998).

On the other hand, evidence is accumulating that many early-type
galaxies in clusters were relatively recently transformed from star
forming galaxies. Dressler et al.\ (1997) report a high fraction of spiral
galaxies in clusters at $0.3< z < 0.5$. These galaxies
are nearly absent in nearby rich clusters (Dressler 1980), and hence
must have transformed into early-type galaxies between $z=0.5$ and
$z=0$. Other studies (e.g., Couch et al.\ 1998, van Dokkum et al. 2000)
have confirmed this trend, and extended it to $z=0.8$. Dressler et al.\
(1997) found that the increased fraction of spiral galaxies at high
redshift was accompanied by a low fraction of S0 galaxies, and concluded that
the $z \approx 0.4$ spiral galaxies transformed into S0 galaxies.
Other studies have found
evidence for merging and interactions in high redshift clusters (Lavery
\& Henry 1988; Lavery, Pierce, \& McClure 1992; Dressler et al.\ 1994;
Couch et al.\ 1998; van Dokkum et al.\ 1999). These transformations may
provide a way to form young elliptical galaxies in the clusters at late
times. The morphological evidence for recent and rapid evolution in
the cluster galaxy population is supported by photometric and
spectroscopic studies, which have found a significant increase of the number
fractions of star forming and post star burst galaxies from $z=0$ to
$z=1$ (e.g., Butcher \& Oemler 1978,
1984, Dressler \& Gunn 1983, Couch \& Sharples 1987, Postman, Lubin, \&
Oke 1998, Poggianti et al.\ 1999, van Dokkum et al 2000). Hence the
evidence appears secure that many early-type galaxies in clusters were
transformed from star forming galaxies at fairly recent times.

The purpose of this paper is to produce models which can bring
these apparently different lines of evidence regarding the ages
into agreement. Specifically, we analyse the effects of late morphological
transformations as found by Dressler et al.,
and others, on the observed
evolution of colors and luminosities with redshift.
We show that the effect of ``progenitor bias''
(Franx \& van Dokkum 1996; van Dokkum et al.\ 2000)
can be very significant: if early-type
galaxies were transformed from other galaxy types at late times, the
youngest progenitors of present-day early-type galaxies drop out of
the sample of early-type galaxies at high redshift.  Hence the samples
of galaxies at low and high redshift are not the same, and in
particular, the high redshift samples might be biased towards the oldest
progenitors of present-day early-type galaxies.
The purpose of this paper is to quantify the effects of progenitor
bias, and to estimate the corrections which are required for models
which ignore morphological evolution.
To that end, we construct relatively simple models for the history of
star formation and morphological type of cluster early-type galaxies.
We show that the observed slow
luminosity evolution and low scatter in the color-magnitude relation and
the Fundamental Plane are natural consequences of recent and ongoing
assembly of early-type galaxies. We correct the observed evolution for
the progenitor bias, and derive the corrected mean age of the stars in
early-type galaxies.

Our models are extensions of the earlier work by van Dokkum et al
(1998b), Bower et al.\ (1998), and Shioya \& Bekki (1998).
These authors considered complex star
formation histories of cluster galaxies, but did not investigate the
effects of morphological transformations on the observed properties of
early-type galaxies at high redshift. The progenitor bias is
implicitly included in the very detailed semi-analytical models by,
for example, Kauffman \& Charlot (1998) and Baugh, Cole, \& Frenk
(1996). In these complex models it is difficult to isolate and quantify
specific effects, such as the progenitor bias.  The main contribution
of our work is that we use simple analytic models with only three free
parameters, which allows us to provide explicit estimates for the
effects of morphological evolution on the observed colors and
luminosities of early-type galaxies.

\section{Modeling}
\label{model.sec}

We construct simple models for the evolution of early-type galaxies,
which parameterize both their star formation history and
morphological evolution. Our models allow the full history to be quantified
by three parameters. They are not based on a full physical description of
the formation and evolution of galaxies as derived by, e.g.,
Kauffmann, White, \& Guiderdoni (1993) and Baugh et al.\ (1996).

\subsection{Basic Description}

We assume that early-type  galaxies have similar histories of star formation
and morphology. The history of each individual galaxy can be described by
three parameters:
the time when star formation starts, the time when
star formation stops,
and the variation of star formation between the start and end.
The star formation is allowed to increase or decrease with time.
It is assumed that the galaxy is classified as an early-type galaxy
some time after the cessation of starformation.
We assume that the galaxies do not stop  their star formation
at the same time, but rather have a distribution of cessation times.

This description is fairly generic for many detailed theories of galaxy 
formation.  We do not define the processes driving the
morphological transformation and the cessation of star formation.
Various processes have been invoked, including
gas stripping of spiral galaxies
(e.g., Gunn \& Gott 1972, Tytler \& Vidal 1978, Abadi, Moore, \& Bower
1999, Kodama \& Smail 2000),
tidal interactions (e.g., Fried 1988, Moore et al.\ 1996), and
mergers (e.g., Lavery et al.\ 1992, van Dokkum et al.\
1999).
It does not matter in our analysis what process causes the cessation
of star formation, and we allow for bursts of star formation to
accompany this event.

The transition phase between star forming galaxy and early-type galaxy
is assumed to last for a significant time.  Immediately after the
cessation of star formation galaxies are likely to show tidal features
(in case of an interaction or merger), or weak spiral arms.  The
morphologies of ``E+A'' or post star burst galaxies seem to be
consistent with this scenario.  They generally are classified as
early-type spirals or tidally distorted systems (Poggianti et al.\
1999; Fabricant et al.\ 2000; van Dokkum et al.\ 2000).  It is not
well known how long it takes before these structures fade or disperse;
we assume that early-type galaxies are first classified as such $\sim
1.5$\,Gyr after star formation in their progenitors ceased.

In the current Section we do not specify whether the models apply to
elliptical galaxies, S0 galaxies, or both. Observationally,
the early-type galaxy fraction is more robust than the relative
numbers of elliptical galaxies and S0 galaxies (e.g., Fabricant,
Franx, \& van Dokkum 2000, van Dokkum et al.\ 2000).
If different histories apply to the two classes of galaxies, two 
separate models will be needed for each sub-class.
We explore such models in Section 5.

\begin{figure*}[t]
\epsfxsize=13cm
\epsffile{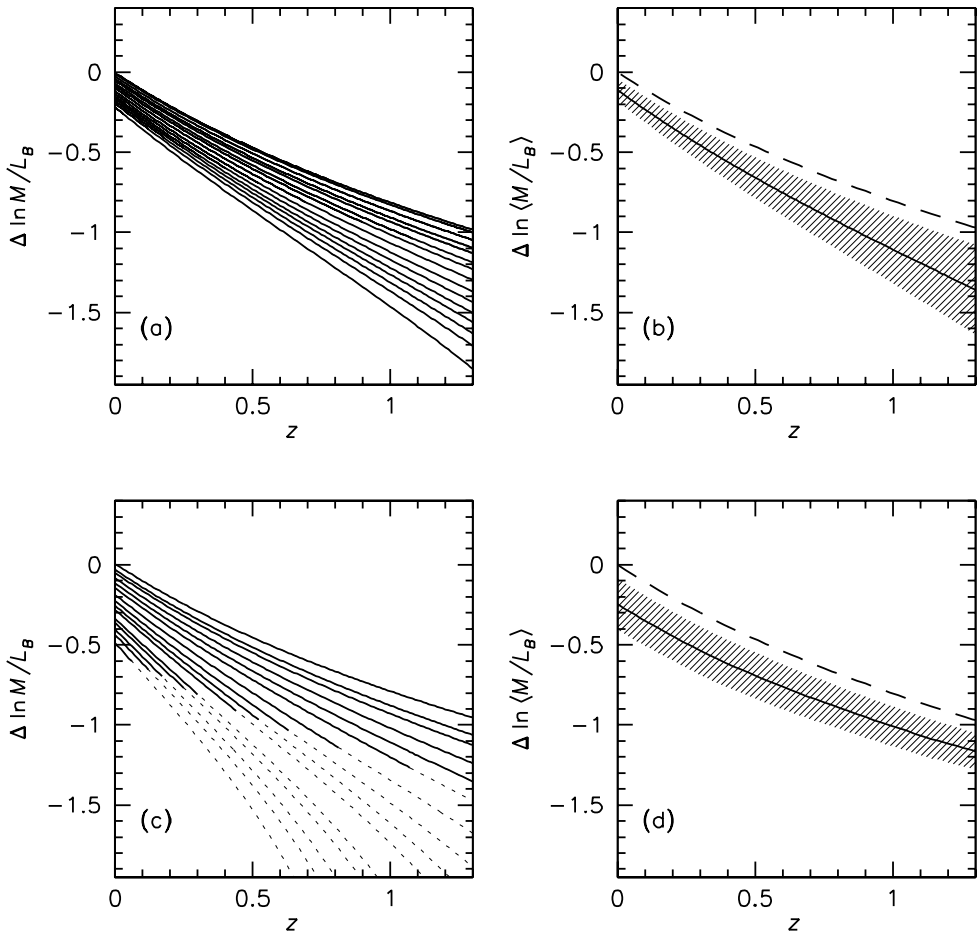}
\caption{\small
Illustration of the evolution of early-type galaxies in traditional
models (a,b) and in models with morphological evolution (c,d). Lines
in (a) and (c) show the evolution of individual galaxies, with a range
of ages. The dashed sections of the curves in (c) indicate that the
galaxies are not yet recognized as early-type galaxies.
Panels (b) and (d) show the evolution of the mean
$M/L$ ratio in these two models. The scatter is indicated by the width
of the hatched regions. The dashed line shows
the evolution of a single age stellar population formed at $z=\infty$.
In models with
morphological evolution the mean $M/L$ ratio evolves slowly and the
scatter is roughly constant, because more and more young galaxies drop
out of the sample at high redshift.
\label{modeltracks.plot}}
\end{figure*}

The generic effects of late morphological transitions are illustrated
in Fig.\ \ref{modeltracks.plot}. 
In panel (a), the evolution of the
mass-to-light ratio is shown for a population of galaxies with a small
range in age, without any late morphological transitions. As can be
seen, galaxies with young stellar populations evolve more rapidly than
galaxies with old stellar populations, and hence the spread in $M/L$
ratios increases with redshift (panel b).  The mean evolution is
faster than that of a single age population formed at $z=
\infty$, which is indicated by the dashed line in (b).
This type of modelling has been used extensively
to constrain the ages of early-type galaxies
in clusters: if there are no morphological transitions, the evolution
of the mean color or luminosity constrains the mean age of early-type
galaxies, and the scatter constrains the spread
in ages (e.g., Bower, Lucey, \& Ellis 1992, Ellis et al.\ 1997,
van Dokkum et al.\ 1998(a,b), Bower et al.\ 1998, Stanford
et al.\ 1998, Ferreras \& Silk 2000).


In panel (c), a model is shown in which galaxies are transformed into
early-type galaxies at a constant rate.  The broken lines indicate
galaxies before they are recognized as early-type galaxies. It can
easily be seen that the sample of early-type galaxies at high redshift
is a small subsample of all galaxies which are classified as
early-type galaxies at $z=0$. The two samples are therefore not
directly comparable, and erroneous results are derived if
morphological evolution is ignored.  As a result, the evolution in the
mean $M/L$ ratio of the early-type galaxies is very slow (panel d).
The slope of the $M/L$ -- $z$ relation is comparable to the slope
for a single stellar population which formed at $z=\infty$ (indicated by the
dashed line), even though the mean formation redshift of the
stars in all
early-type galaxies at $z=0$ is low at $\langle z_* \rangle \approx 2$. Even
more remarkable is the fact that the scatter in $M/L$ ratios is virtually
constant. These effects are caused by the fact that the youngest galaxies
continuously drop out of the sample going to higher redshifts.
In the subsections below we specify a broader range
of models, and explore the consequences.

\subsection{Model Parameters and Methodology}

The full models are quantified by three parameters: $\tstart$,
the time when star formation starts, $\taustop$,
the time scale which characterizes the distribution of times when
star formation
stops, and $\fstar$, which describes the star formation rate between
the start and end of star formation.
The last two parameters are defined in the following way.

The parameter $\taustop$ determines the probability distribution of
$\tstop$, the time when star formation stops for an individual
galaxy:
\begin{equation}
\label{first.eq}
P(\tstop) \propto \exp \left( - \frac{t}{\taustop}\right).
\end{equation}
We show later that this expression can provide a satisfactory fit to the
data. 
If $\taustop \ll \t0$, with $\t0$ the present age
of the Universe, star formation in the progenitors
terminated at very high redshift. At the other extreme,
$\taustop=\infty$ corresponds to a constant transformation
rate. 
For each individual galaxy, the morphological transformation to early-type
galaxy occurs at $\tstop + 0.1 \t0$,
i.e., $\sim 1.5$ Gyr after truncation of star formation.

The parameter $\fstar$ characterizes the variation of
the star formation rate.
The star formation history of the  galaxies 
can include bursts, and increasing or decreasing
continuous formation rates as a function of time.
After star formation ceases, the luminosity and color evolution of such a
complex population is well approximated by a single age population of stars
with the same luminosity weighted mean age (e.g., van Dokkum et al.\ 1998b).

Hence for each galaxy we approximate the evolution  of the complex population
by that of a single age population formed at $t=t_*$, with
\begin{equation}
t_* = \fstar \tstop + (1-\fstar) \tstart.
\label{tform.eq}
\end{equation}
For $\fstar\approx 1$, the stellar population is dominated by a burst at the
end of the star formation history, and for $\fstar \approx 0$ the population
is dominated by a burst at the start of the star formation
history. If the
star formation rate is approximately constant from $t=\tstart$
to $t=\tstop$, then $\fstar \approx 0.5$.

We tested the accuracy of our approach by calculating the luminosity
and color evolution of galaxies with complex star formation histories
(e.g., an exponentially declining star formation rate followed
by a star burst), and comparing the results
to predictions from single burst models with the same values
of $f_*$. After transformation to early-type galaxy (i.e., $>1.5$\,Gyr
after star formation has ceased) our approximation of
luminosity and color evolution is accurate to a few percent.
Three possible star formation
histories of early-type galaxies
are shown in Fig.\ \ref{fstar.plot}. The model with $f_*=0.7$ has
a star burst at the end of its star formation history, and may
be appropriate for spiral galaxies falling into clusters,
or mergers. Although these histories are quite
complex, $\gtrsim 1.5$\,Gyr after star formation has ceased
their evolution is similar to that of
single age stellar populations formed at $t_* = \fstar \tstop
+ (1-\fstar) \tstart$.


\vbox{
\begin{center}
\leavevmode
\hbox{%
\epsfxsize=7.5cm
\epsffile{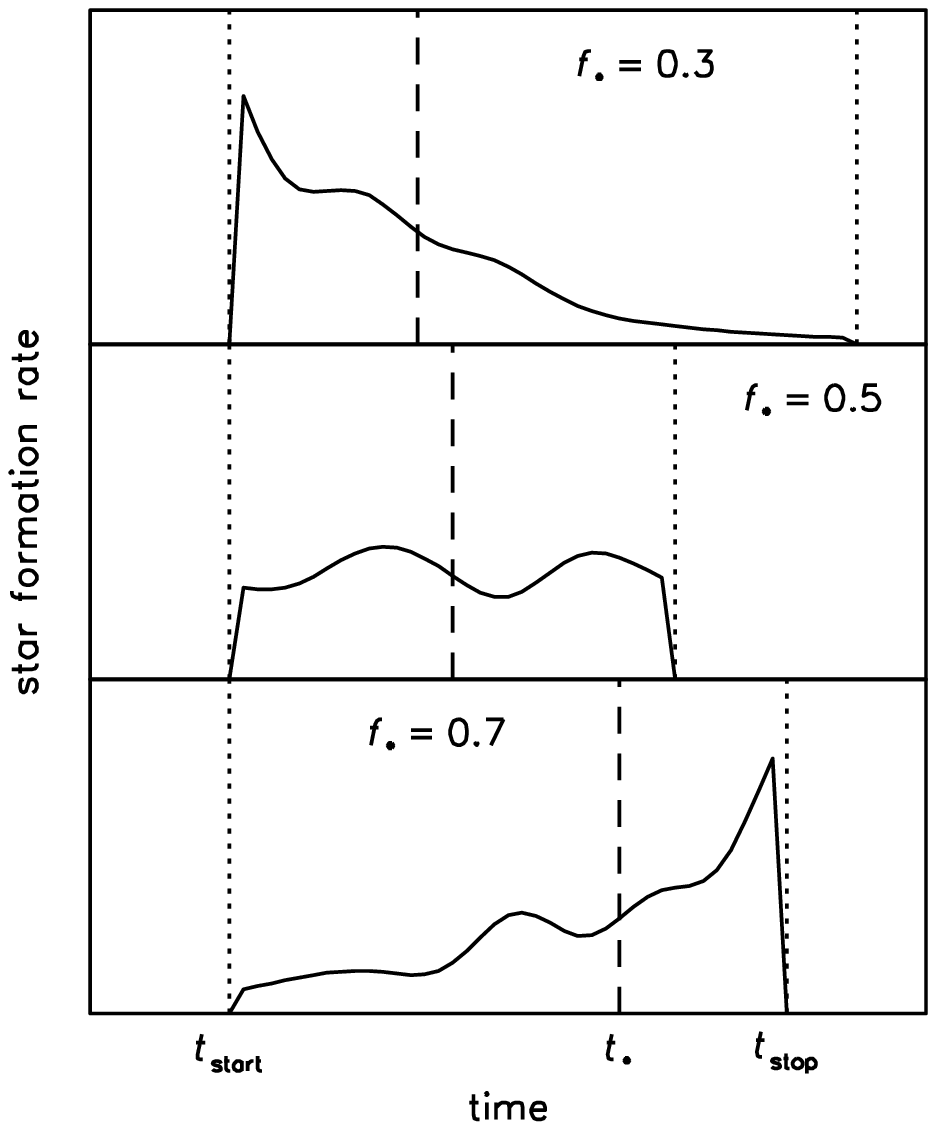}}
\figcaption{\small
Illustration of our parameterization. Galaxies start forming stars
at $\tstart$, and star formation terminates at $\tstop$. Galaxies
do not stop their star formation at the same time, but have a distribution
of cessation times, parameterized by the transformation timescale
$\taustop$. The star
formation history between $\tstart$ and $\tstop$ can be quite complex,
and may include star bursts. It is parameterized by the parameter
$f_*$, which describes whether star formation is more weighted towards
$\tstart$ or $\tstop$.
\label{fstar.plot}}
\end{center}}

Our parameterization allows straightforward computation of
luminosities and colors of galaxies, because
the luminosity evolution of a single age stellar population can
be approximated by a power law (e.g., Tinsley 1980, Worthey 1994).
Therefore, to good approximation,
\begin{equation}
\label{levo.eq}
L \propto \frac{1}{(t - t_*)^{\kappa}}.
\end{equation}
The coefficient $\kappa$ depends on the passband, the IMF,
and the metallicity. In this paper we will limit the discussion
to rest frame $B$ band luminosities, and rest frame $U-B$ colors.
Our results can easily be expressed in other (rest frame) bands.
The Worthey (1994) models give
$\kappa_B = 0.91$ and $\kappa_U=1.07$ for solar metallicity and
a Salpeter (1955) IMF. The color evolution can be
approximated by
\begin{equation}
\label{colevo.eq}
\frac{L_B}{L_U} \propto (t-t_*)^{\kappa_U - \kappa_B}.
\end{equation}
Expressed in magnitudes, color and luminosity evolution
are related through
\begin{eqnarray}
\Delta (U-B)& =& (\kappa_U - \kappa_B) \kappa_B^{-1} \Delta B \\
            & \approx & 0.18 \Delta B
\label{agecol.eq}
\end{eqnarray}
in these models.
The evolution of the mean color and scatter in the color will
therefore be similar to the scaled evolution of the mean $M/L$ ratio and
its scatter.

In the following,
we use Monte-Carlo simulations to calculate model predictions for
given values of $\taustop$, $\fstar$, and $\tstart$.
This approach has the advantage that the mean and scatter in
colors and luminosities at a given time can be
computed using the same methods as for the observations.
In the simulations, all galaxies start out
as star forming objects at $t=\tstart$. Each galaxy is assigned
a value for $\tstop$, the time when star formation ceases.
The distribution of $\tstop$ is given by Eq.\ \ref{first.eq},
with boundary condition $0<\tstop<0.9 \t0$, with $\t0$ the
present age of the Universe.
At each timestep $t$, objects which satisfy the condition $\tstop
+0.1 \t0 < t$ are selected as early-type galaxies.  Using Eq.\
\ref{levo.eq} and \ref{colevo.eq} these galaxies are assigned a
luminosity and a color.  The central value and the spread of the color
and luminosity distribution at time $t$ are calculated with the
biweight statistic (Beers, Flynn, \& Gebhardt 1990). This statistic
was also used by Stanford et al.\ (1998) and van Dokkum et al.\
(1998b, 2000) to calculate the scatter in the high redshift
color-magnitude relation.
We note that
the evolution of the mean $M/L$ ratio and colors of early-type
galaxies can also be evaluated analytically.
In Appendix A we give approximate
expressions which can be used for most applications.

\subsection{Examples of Models with Progenitor Bias}

\begin{figure*}[t]
\epsfxsize=17cm
\epsffile{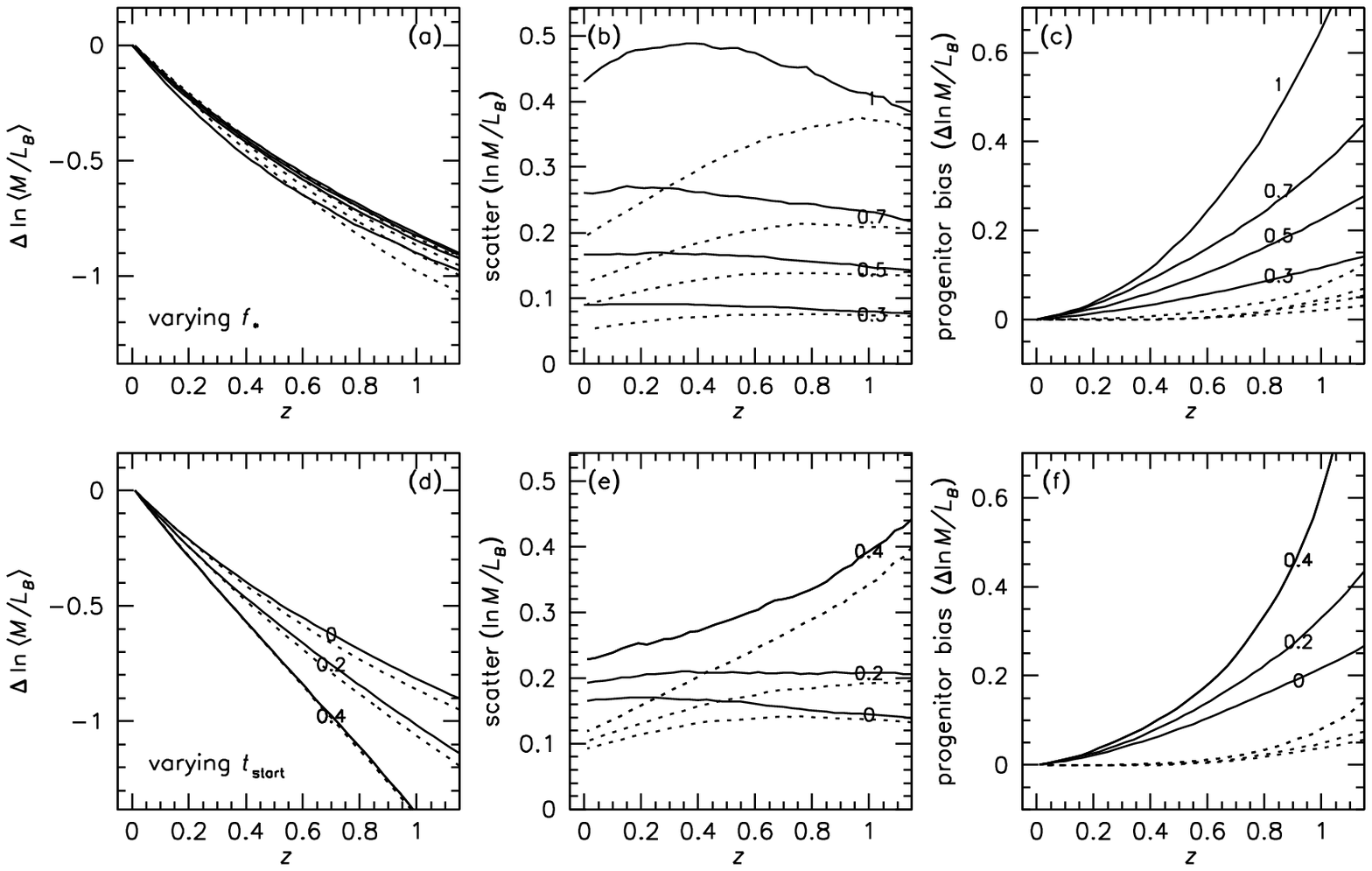}
\caption{\small
Predicted evolution of the mean $M/L$ ratio and its scatter in
various models with progenitor bias. 
In all panels, solid lines show models
with strong morphological evolution to $z=1$ ($\taustop=0.5 \t0$)
and broken lines show models with mild morphological evolution
($\taustop=
0.2 \t0$). Panels (a,b,c) demonstrate the effect of varying the
star formation history, parameterized by $\fstar$. The evolution of
the mean $M/L$ ratio (panel a) is remarkably similar in all models, whereas
the scatter (panel b) and
the progenitor bias (panel c) are very sensitive to the star formation
history. Panels (d,e,f) show the effect of varying the time of onset
of star formation, $\tstart$. Varying $\tstart$ has a large effect
on the evolution of the mean $M/L$ ratio, as well as on the
scatter and the progenitor bias. The models indicate that
the observed transformation rate and the observed scatter suffice
to predict the progenitor bias.
\label{modeldata.plot}}
\end{figure*}

We calculate the predicted evolution of the $M/L$ ratio and its scatter
for several models. 
To demonstrate the effects of progenitor
bias we consider two classes of models, one with mild morphological evolution
($\taustop=0.2 \t0$) and one with strong morphological evolution
($\taustop=0.5 \t0$).  Figure \ref{modeldata.plot}
shows the results.


In models with $\taustop=0.2 \t0$ morphological
evolution is not very important to $z=1$, and $\approx 80$\,\% of present-day
early-type galaxies were already in place at that redshift. In models
with $\taustop=0.5 \t0$, on the other hand, only $\approx 50$\,\%
of early-type galaxies were in place at $z=1$.
In all panels of Fig.\ \ref{modeldata.plot} 
broken lines show the class of
models with mild morphological evolution, and solid lines show
models with strong morphological
evolution. 

\subsubsection{Effect of Varying the Star Formation History}
\label{progz.sec}

Panels (a), (b), and (c) show the effect of changing the star
formation history, while keeping the time of onset of star formation
fixed at $\tstart = 0$.  The results for the evolution of the mean
$M/L$ ratio are shown in panel (a).  Quite remarkably, nearly all
model predictions lie very close to the line for old populations which
formed at the start of the Universe ($z_{\rm form}=\infty$). We find
little dependence on the tranformation time scale $\taustop$ or on
$f_*$, which characterizes the star formation history.  This result
illustrates the dramatic effects of the progenitor bias, as the mean
stellar age of the galaxies at $z=0$ can be as low as half the age of
the Universe (for $f_* = 1$), and yet the apparent evolution follows
the relation for very old populations formed at infinite redshift. The
reason for this behaviour is that the formation models are almost
scale free.  In models with a constant transformation rate, at any
time the youngest half of the galaxies have been added to the sample
in the last half of the age of the Universe at that time.  As a
result, the mean stellar age of the galaxies is an almost constant
fraction of the age of the Universe, independent of the star formation
histories of individual galaxies, and the mean mass-to-light ratios
and colors will evolve as if the stars formed at very high redshift.
In the Appendix we demonstrate this effect analytically.  The absolute
colors and mass-to-light ratios at $z=0$ will be different for these
models, but these are notoriously difficult to interpret uniquely
(e.g., Worthey 1994, Faber et al.\ 1999).

Fortunately, the scatter in $M/L$ ratios and colors is much more
sensitive to the transformation timescale and star formation
histories.  We show the scatter in $M/L$ ratios in panel (b).  For a
given morphological transformation rate the variation in the average
scatter is large, because it is a strong function of the star
formation history of galaxies.  As we will see later, this implies
that observations of the scatter put important constraints on the
models.  For galaxies with later star formation, $f_*$ increases, and
the scatter increases almost linearly with $f_*$.  For the limiting
case of a constant transformation rate ($\taustop=\infty$), we find
that the scatter is well approximated by $\sigma (\ln M/L_B) \sim 0.3
f_*$ (see Appendix).  The scatter is not strongly dependent on
redshift because of the scale-free nature of the models. In such
scale-free models, the relative age differences between early-type
galaxies at any epoch is fairly constant. The scatter is proportional
to the age differences between galaxies relative to the mean age
(e.g., van Dokkum et al.\ 1998b). As galaxies are continuously added
to the sample, the range in ages increases at the same rate as the
mean age; hence the spread in ages relative to the mean age remains
constant. In the Appendix this effect is demonstrated
analytically. Only in models with mild morphological evolution to
$z=1$ (indicated by the broken lines in Fig.\ \ref{modeldata.plot}) we
find an increase of the scatter with redshift. In these
models the scatter is very low at $z=0$, and at $z>0.5$ reaches values
similar to those in models with strong morphological evolution.

The progenitor bias of the models is shown in panel (c). We define the
progenitor bias at given redshift as the difference between the $M/L$
ratio of early-type galaxies and the $M/L$ ratio of
all progenitors of present-day early-type galaxies.  This is the
``error'' in the observed $M/L$ ratio that is caused by the late
addition of early-type galaxies to the sample.
As the figure shows, the bias increases with increasing tranformation
time scale $\taustop$, and increasing $f_*$. Both parameters also cause the
scatter to increase.
For models with strong morphological evolution the progenitor bias
can be approximated by
\begin{equation}
\label{progz.eq}
{\rm bias}(\ln M/L_B) \approx 1.3 \times {\rm scatter}(\ln M/L_B) \times z
\end{equation}
This approximation is accurate to $\lesssim 10$\,\% for $\taustop>0.5 \t0$.
This result suggests that the progenitor bias can be estimated on the basis
of the  observed transformation rate, and the observed scatter.
The star formation history is the individual galaxies, as parametrized
by $f_*$, is not needed to estimate the effect.
This is a very useful result, as it is difficult to constrain the
value of $f_*$ directly from the observed colors and luminosities.

\subsubsection{Effect of Varying the Time of Onset of Star Formation}

Panels (d), (e), and (f) show the effect of changing the time of
onset of star formation $\tstart$, while keeping the star formation
history constant at $f_*=0.5$ (i.e., approximately
constant star formation from $\tstart$ to $\tstop$). The evolution
of the mean $M/L$ ratio for the various models is shown in panel (f).
The evolution is very sensitive to the time of onset of star
formation: there is an almost linear relation between the
time when star formation commences and the rate of $M/L$ evolution.
The reason for this behaviour is that the mean age of the stellar
population in all galaxies is lower for higher values of
$\tstart$. As will be shown in Sect.\ \ref{mc.sec} the observed evolution
of the mean $M/L$ ratio places strong constraints on the time
when star formation commenced in early-type galaxies.

The scatter and the progenitor bias are shown in panels (e) and
(f). They both depend on the value of $\tstart$, such that the scatter
is higher and the progenitor bias stronger for later onset of star
formation. As a result of this dual dependence
the relation between the progenitor bias
and the observed scatter (Eq.\ \ref{progz.eq}) is not very
sensitive to the value of $\tstart$, once again indicating
that the observed rate of morphological evolution and
the observed scatter suffice to estimate the progenitor bias.

\section{Application to Observations}
\label{mc.sec}

The models described in the previous Section can be applied
to observations of early-type galaxies in clusters at $0<z<1$.
Parameters to fit are the evolution of {\em (i)} the early-type
galaxy fraction, {\em (ii)} the mean $M/L_B$ ratio, as determined
from the Fundamental Plane, {\em (iii)} the scatter in $M/L_B$,
and {\em (iv)} the scatter in the color-magnitude
relation.
Free parameters are the morphological transformation rate
(parameterized by $\taustop$), and
the star formation histories of
galaxies prior to the transformations (parameterized by $\fstar$
and $\tstart$).

\subsection{Data}


\begin{figure*}[t]
\epsfxsize=18cm
\epsffile{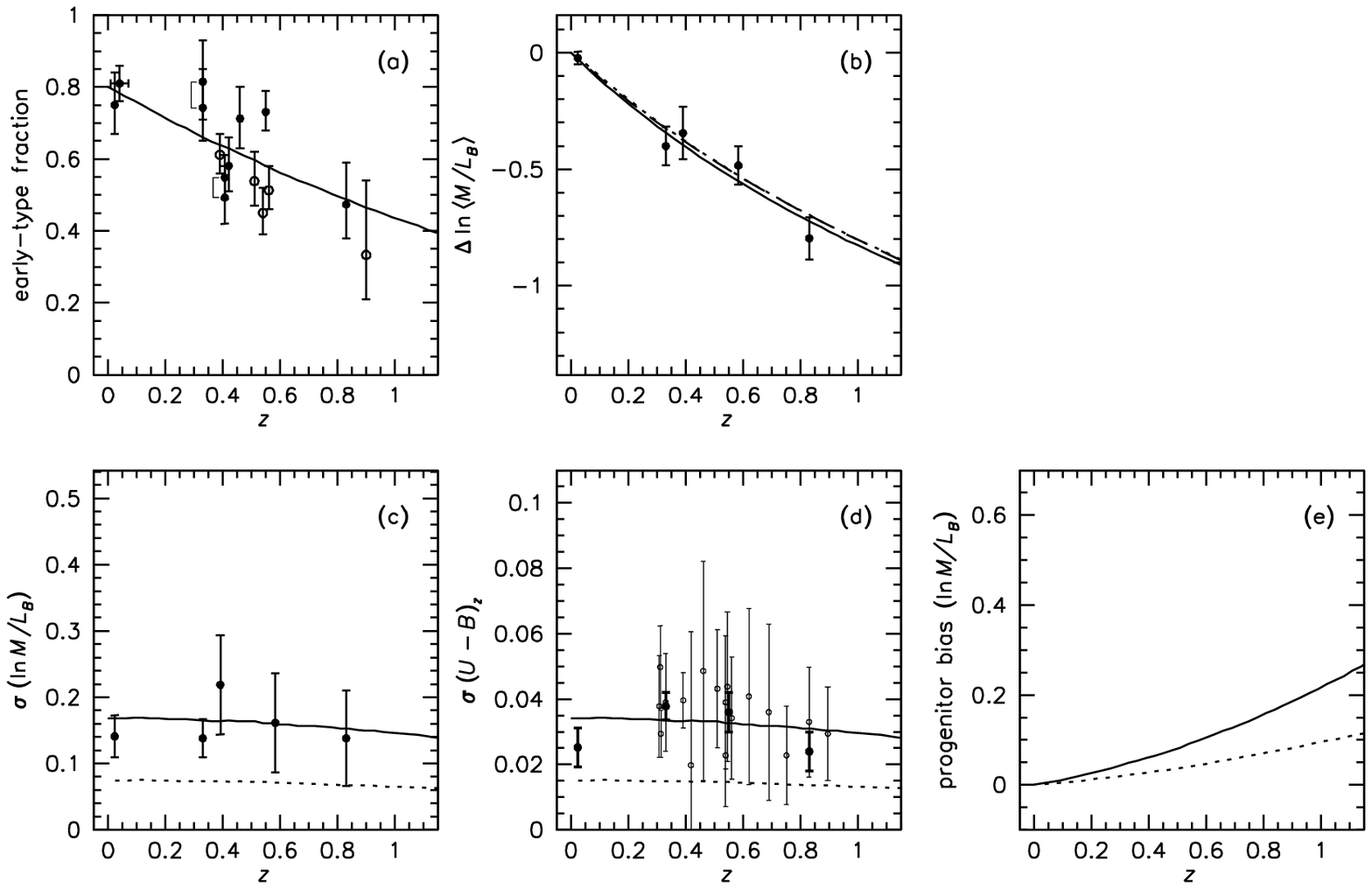}
\caption{\small
Comparison of observations to model predictions.
Shown are
the evolution of the early-type galaxy fraction (a), the evolution
of the mean $M/L_B$ ratio (b), the scatter in $\ln (M/L_B)$, as
derived from the scatter in the Fundamental Plane of early-type
galaxies (c), and the scatter in the $U-B$ color-magnitude relation
(d).  Sources of the data are described in the text. The solid lines
show the best fitting model, with 
$\taustop = 0.5 \t0$, $\fstar=0.5$ (i.e., a constant star formation
rate), and $\tstart =0$. This model
provides an excellent fit to the data. The broken lines show a model
with $\fstar=0.25$ (i.e., a declining star formation rate).
This model allows for additional scatter in the
FP and the CM relation as a result of, e.g., metallicity variations
or mergers.
The progenitor bias is shown in panel (e). It is a strong
function of $f_*$, and hence the scatter. The observed scatter
provides an upper limit to the bias of $\approx 0.2$ in $\ln (M/L_B)$ at
$z=1$.
\label{bestfit.plot}}
\end{figure*}

The evolution of the early-type galaxy fraction in clusters is shown
in Fig.\ \ref{bestfit.plot}(a) (taken from van Dokkum et al.\ 2000).
Data points are from Dressler (1980), Andreon, Davoust, \& Heim
(1997), Dressler et al.\ (1997), Lubin et al.\ (1998), Fabricant et
al.\ (2000), and van Dokkum et al.\ (2000). The early-type fraction
decreases by a factor $\sim 2$ from $z=0$ to $z=1$.  The scatter
around the downward trend is significant, and it will be interesting
in the future to explore systematic differences in the evolution
depending on cluster type. In this Section we do not distinguish
elliptical galaxies and S0 galaxies within the class of early-type
galaxies.  In Sect.\ \ref{ES0.sec} we explore models in which both
classes are modeled separately.

The evolution of the early-type galaxy fraction is possibly influenced
by the inclusion of low mass galaxies undergoing star bursts. Since
the samples are magnitude selected, the presence of such galaxies at
high redshift would decrease the fraction of early-type galaxies, and
would be unrelated to the evolution of massive galaxies.  For the
cluster \1054{} at $z=0.83$ we have a large data set of confirmed
members with accurate colors (van Dokkum et al.\ 2000), and we tested
the importance of this effect by determining the early-type galaxy
fraction among red galaxies alone. If we limit the analysis to red
galaxies with $(U-B)_z >0.3$ the early-type fraction changes from
45\,\% to 52\,\%.  This result indicates that the brightening of low
mass galaxies has a minor effect on the fraction of early-type
galaxies.  Other effects may have the opposite effect. As an example,
biases introduced by the selection of the clusters themselves probably
cause us to {\em under}estimate the evolution of the early-type galaxy
fraction (see, e.g., Kauffmann 1995).

The evolution of the rest frame $M/L_B$ ratio with redshift is shown
in Fig.\ \ref{bestfit.plot}(b) and is taken from van Dokkum et al.\
(1998a). Data are from J\o{}rgensen et al.\ (1996), van Dokkum \&
Franx (1996), Kelson et al.\ (1997), and van Dokkum et al.\
(1998a). The $M/L$ ratio evolution is derived from the evolution of
the zeropoint of the Fundamental Plane relation (see van Dokkum \&
Franx 1996). The evolution is well determined, because the
Fundamental Plane has very small scatter.

The scatter in $\ln (M/L_B)$ is shown in Fig.\ \ref{bestfit.plot}(c).
The data point at $z\approx 0$ is from J\o{}rgensen et al.\ (1996).
Data points at higher redshift are from
Kelson et al.\ (2000) ($z=0.33$), Kelson et al.\ (1997)
($z=0.58$), and van Dokkum et al.\ (1998a) ($z=0.83$).
The highest redshift points have considerable uncertainty,
because they are derived from small samples.

The scatter in the $U-B$ color-magnitude relation is
taken from van Dokkum et al.\ (2000) and
shown in Fig.\ \ref{bestfit.plot}(d). Data are
from Stanford et al.\ (1998), Bower, Lucey,
\& Ellis (1992), Ellis et al.\ (1997), van Dokkum et al.\ (1998b),
and van Dokkum et al.\ (2000). The observations were brought
to a common (rest frame) band by using $\sigma (U-B) = 1.4 \sigma (B-V)$
and $\sigma (U-B) = 0.6 \sigma (U-V)$, as derived from the
Worthey (1994) models. The scatter is roughly constant with
redshift, at $\sigma (U-B) \approx 0.03$ magnitudes.

\subsection{Constraints on Model Parameters}

\subsubsection{Transformation Time Scale}

The observed evolution of the early-type galaxy fraction directly
constrains the transformation time scale $\taustop$. This is
demonstrated in Fig.\ \ref{modeldata.plot}(a), which shows the
redshift dependence of the transformed galaxy fraction for different
values of the transformation time scale.

We varied the transformation time scale between $\taustop = 0$ and
$\taustop=\infty$. If $\taustop=0$
all morphological transformations occur at very high redshift and
if $\taustop=\infty$ the transformation rate is constant.
The observed evolution of the early-type galaxy fraction, shown
in Fig.\ \ref{bestfit.plot}(a), gives
a best fitting value of $\taustop \approx 0.5 \t0$. Values between
$0.3 \t0$ and $1.7 \t0$ are consistent with the data
at the 90\,\% confidence level.

The clusters in Fig.\ \ref{bestfit.plot}(a) form a very heterogeneous
sample, and a potential worry is that the evolution is severely
affected by selection biases. As a test of the robustness of the
transformation rate we fitted all clusters with high X-ray
luminosities separately. Solid symbols in Fig.\ \ref{bestfit.plot}(a)
are clusters with $L_X>10^{44.5}\,h_{50}^{-2}$\,ergs\,s$^{-1}$ (see
van Dokkum et al.\ 2000). This sample gives a very similar best fit
value of $\taustop \approx 0.45 \t0$. We note, however, that the
morphological transformation rate is one of the main uncertainties
in our study. We will return to this issue in Sect.\ \ref{ES0.sec}.

\subsubsection{Star Formation Histories}

We next explore models with the best fit transformation time
$\taustop=0.5 \t0$, and we vary the time of onset of star formation
$\tstart$ and the parameter describing the star formation history
$\fstar$.

As demonstrated in the previous Section, the rate of
evolution of the mean
$M/L$ ratio has an approximately linear dependence on the
value of $\tstart$. The observed evolution, shown in
Fig.\ \ref{bestfit.plot}(b), is very slow. The dashed line indicates
a simple single burst model with $z_{\rm form} = \infty$. This model
provides a good fit to the data, as was previously shown by
van Dokkum et al.\ (1998a). We find a best fitting value of
$\tstart = 0.03 \t0$. Values between $\tstart = 0$ and $\tstart = 0.2\t0$
are consistent with the data at the $90$\,\% confidence level.
The implication is that star formation commenced at high
redshift ($z>2.5$) in early-type galaxies. The line in Fig.\
\ref{bestfit.plot}(b) shows the best fitting model with $\tstart=0$.

As was demonstrated in Sect.\ \ref{model.sec}
the star formation history between $\tstart$ and $\tstop$ is
constrained by the observed scatter in $M/L$ ratio and color.  This
strong dependence of the scatter on the star formation history was
discussed earlier by van
Dokkum et al.\ (1998b) and Bower et al.\ (1998) in the context of
``traditional'' models which do not include morphological evolution.
If all scatter in the color-magnitude relation and the Fundamental Plane
is produced by age variations, the best fitting model
has $\fstar = 0.52$ (i.e., an approximately
constant star formation rate), with values to
$\fstar=0.62$ consistent with the data at the 90\,\% confidence
level. The lines in Fig.\ \ref{bestfit.plot}(c) and (d) show the
model with $\fstar=0.5$.
For values of $\fstar$ lower than
0.5 the scatter is underpredicted, and additional scatter due to,
e.g., merging, metallicity variations, or dust is allowed.  As an
example, the broken lines show a model with $\taustop=0.5 \t0$,
$\tstart = 0$, and $\fstar = 0.25$.  The preferred values of $\fstar
\leq 0.5$ correspond to gradual
star formation histories similar to
those depicted in the upper two panels of Fig.\ \ref{fstar.plot}.

\subsection{The Progenitor Bias}

Our best fitting model has $\taustop = 0.5 \t0$, $\t0 = 0$, and
$\fstar \leq 0.5$.
The progenitor bias,
expressed as the difference between the observed $M/L$ ratio
of early-type galaxies and the $M/L$ ratio of all progenitors of
early-type galaxies in nearby clusters, is shown in Fig.\
\ref{bestfit.plot}(e).
The progenitor bias is maximal if $\fstar=0.5$, in which case the
scatter in the CM relation and the FP
is entirely produced by age variations.
In this extreme model the progenitor
bias is approximately $0.2 \times
z$ in $\ln M/L_B$, or $0.04 \times z$ in $U-B$ color.

We stress that these values are upper limits, because the
scatter can be caused by other effects than age. As shown in
the previous Section, the progenitor bias scales linearly with the
scatter introduced by age variations. For example, if $\fstar = 0.25$
the scatter due to age is only half the observed scatter.  In this
case the progenitor bias is only $0.1 \times z$ in $\ln M/L_B$
(indicated by the broken line in Fig.\ \ref{bestfit.plot}e).

It is interesting to
compare these results to the simple scaling with the observed scatter
that was derived in Sect.\ \ref{progz.sec}. Using the observed scatter
of $\approx 0.15$ in $\ln M/L_B$ Eq.\ \ref{progz.sec} gives a
progenitor bias of $\sim 0.2 z$, identical to the value we derived
from the full fitting of the models to the observations. This
remarkable agreement demonstrates quantitatively that the observed
transformation rate and the observed scatter suffice to predict the
progenitor bias. In particular, knowledge of the detailed star
formation histories of the galaxies ($\tstart$ and $\fstar$ in our
models) is not required.

\subsection{Dependence on Cosmology}

We used $\Omega_m=0.3$ and $\Omega_{\Lambda}=0$ to calculate
the model predictions and the data points shown
in Fig.\ \ref{bestfit.plot}. The models depend on the
cosmological model because of the conversion from time to
redshift. The observed evolution of the mean $M/L$ ratio also depends
on the cosmology, because effective radii of galaxies are measured in
arcseconds. As demonstrated in Bender et al.\ (1998) and
van Dokkum et al.\ (1998a) the
dependencies can be quite strong, and the observed evolution of $\ln
M/L_B$ provides an upper limit on $\Omega_m$ (assuming the slope of
the IMF is not significantly steeper than the Salpeter (1955)
value). Note that the models and data shown in Fig.\
\ref{bestfit.plot} are not dependent on the Hubble constant, because
the luminosity evolution of stellar populations can be approximated by
a power law and all data points are relative to nearby clusters.

Our results are very similar for a Universe with a
cosmological constant ($\Omega_m=0.3$ and $\Omega_{\Lambda}=0.7$).
The predicted evolution of the early-type galaxy fraction remains
virtually unaffected, and the best fitting value for the
transformation time scale remains $\taustop \approx 0.5 \t0$.
The observed evolution of the mean $M/L_B$ ratio is more rapid in this
cosmology, and is no longer well fitted by models with
extremely high formation redshift of the stellar population
(see also van Dokkum et al.\ 1998a). Models with
$\tstart=0$ predict too slow evolution, and good fits
are obtained for $\tstart = 0.15 \t0$.
Because the stellar populations of early-type galaxies are
younger for $\tstart>0$ the predicted scatter in
$M/L$ ratios and colors is higher for a given value of
$\fstar$. As a result, the observed scatter is best fitted for
$\fstar \approx 0.4$ in this cosmology, as compared
to $f_* \approx 0.5$ for a Universe without cosmological constant.

\section{Correcting for Progenitor Bias}

The evolution of
the $M/L$ ratio can be corrected for progenitor bias.
Our best fitting model indicates that the observed
evolution of the $M/L$ ratio underestimates the true evolution by
$\approx 0.2 \times z$ in $\Delta \ln \langle M/L_B \rangle$, and $\approx
0.04 \times z$ in $\Delta \langle U-B \rangle$.
van Dokkum et al.\ (1998b) found $\Delta \ln
\langle M/L_B\rangle = (-0.9 \pm 0.1) z$ for
$\Omega_m=0.3$ and $\Omega_{\Lambda}=0$, or $\Delta \ln \langle M/L_B
\rangle = (-1.1 \pm 0.1) z$ for $\Omega_m=0.3$ and
$\Omega_{\Lambda}=0.7$. The evolution after
applying the maximum correction for progenitor bias is $\Delta \ln
\langle M/L_B \rangle
\approx -1.1 z$, or $\Delta \ln \langle M/L_B \rangle
\approx -1.3 z$ for positive cosmological constant.
We conclude that the progenitor bias has
a small ($\lesssim 20$\,\%), but non-negligible effect on the
observed evolution of the mean $M/L$ ratio from
$z=0$ to $z=1$. The corrections are upper limits, because it is assumed
that the scatter in luminosities and colors is entirely due
to age variations among the galaxies.

The observed evolution of the mean color and $M/L$ ratio of early-type
galaxies indicates a very high redshift of formation for their stars
(e.g., van Dokkum \& Franx 1996, Schade, Barrientos, \& Lopez-Cruz
1997, Stanford et al.\
1998, van Dokkum et al.\ 1998a, Kodama et al.\ 1998). Because the true
evolution of early-type galaxies is underestimated as a result of the
progenitor bias, the ages of their stars are {\em over}estimated.

We constrain the formation epoch of the stars in early-type galaxies
from the corrected evolution of the mean $M/L$
ratio. First we assume that the scatter in the color-magnitude relation and
the Fundamental Plane is entirely caused by age variations.
Using $\Delta \ln \langle L_B \rangle = 1.1 z$ in Eq.\
\ref{levo.eq} we find that the mean luminosity weighted formation
time of the stars in early-type galaxies
$\langle t_* \rangle = 0.16 \t0$, or
$\langle z_* \rangle = 3.0^{+0.9}_{-0.5}$ when expressed in redshift.
For $\Omega_m=0.3$ and
$\Omega_{\Lambda}=0.7$ $\Delta \ln \langle L_B \rangle = 1.3 z$,
and we find $\langle z_* \rangle = 2.0^{+0.3}_{-0.2}$. As a check on our
procedure, we also computed the mean formation epoch directly
from the Monte-Carlo simulation which gave the best fit to the
observations.  The age derived from the simulation is identical
to the age derived from the corrected evolution of the $M/L$ ratio.

If the scatter in the color-magnitude relation and the
Fundamental Plane is not entirely caused by age variations,
the mean age of the stars is higher. Both the mean age and
the scatter have an approximately linear dependence on $\fstar$
(Eq.\ \ref{tform.eq} and \ref{scatevo.eq}). Therefore,
for a given morphological transformation rate
the mean age of the stars is to good approximation a linear
function of the scatter at $z=0$. From the Monte-Carlo
simulations with $\taustop=0.5 \t0$ we find
\begin{eqnarray}
\label{lagescat.eq}
\langle t_* \rangle / \t0 & \sim & 2 \times \sigma_{\ln M/L(B)}\\
 & \sim & 10 \times \sigma_{U-B}
\end{eqnarray}
for $\Omega_m=0.3$ and $\Omega_{\Lambda}=0.7$.
If the contribution of age variations to the scatter is negligible
the correction for progenitor bias is negligible as well, and
we find $\langle z_* \rangle \approx 6.5$ for $\Omega_m=0.3$
and $\Omega_{\Lambda}=0$, and $\langle z_* \rangle \approx 2.6$
for $\Omega_m=0.3$ and $\Omega_{\Lambda}=0.7$
(cf.\ van Dokkum et al.\ 1998a).
Unfortunately it is observationally difficult
to disentangle the contributions of age and metallicity to the
observed scatter (e.g., Trager et al.\ 2000).


\section{Separating Elliptical and S0 Galaxies}
\label{ES0.sec}

In the previous Sections we have treated early-type galaxies as one
class.  However, Dressler et al.\ (1997) separated the evolution of
elliptical and S0 galaxies, and found that the ratio of the fractions
of S0 and elliptical galaxies evolves
rapidly from $z=0$ to $z\approx 0.55$.

Figure \ref{dressler.plot} shows the evolution of the number fractions
of S0 galaxies and elliptical galaxies separately, for the Dressler et
al.\ (1997) clusters (solid symbols).  The dotted lines are linear
fits to the data. As can be seen, the S0 fraction increases by a
factor of $\sim$ 5 towards lower redshift, whereas the elliptical
fraction drops by $\sim 30$\,\% from redshift $z=0.55$ to $z=0$.
Dressler et al.\ (1997) interpret the rapid evolution of the S0
fraction as evidence for transformation of spiral galaxies to S0
galaxies since $z=0.55$. As pointed out by Fabricant et al.\ (2000)
the decrease in the elliptical galaxy fraction over time requires
substantial accretion onto clusters since $z=0.55$.

We model this evolution in the following way: we assume that
the fraction of ellipticals is constant with time, and that S0 
galaxies are formed
from the spiral galaxies that are present
in the clusters at $z=0.55$. We ignore the fact that
infalling spiral galaxies are also transformed into S0 galaxies.
The evolution of the S0 galaxies is approximated reasonably well by
a model with a constant transformation rate (Fig.\ \ref{dressler.plot}).
The overprediction
of the fraction of S0 galaxies at high redshift implies that we underestimate
the differences between the elliptical and S0 galaxies.


\vbox{
\begin{center}
\leavevmode
\hbox{%
\epsfxsize=6.5cm
\epsffile{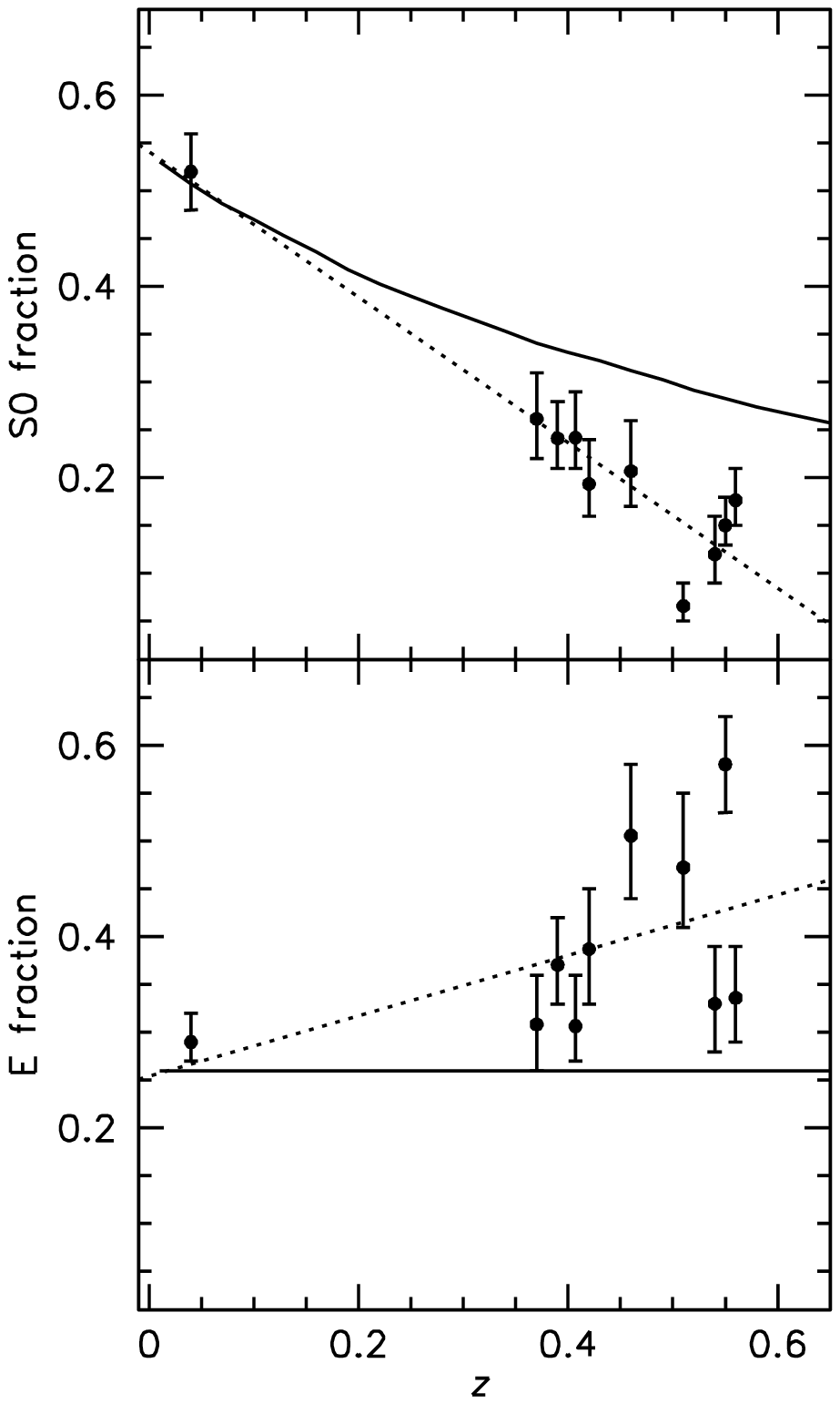}}
\figcaption{\small
Evolution of the number fraction
of elliptical galaxies and S0 galaxies separately,
from Dressler et al.\ (1997) (solid symbols).
The S0 fraction decreases with redshift, whereas the elliptical
fraction increases. The dotted lines are linear fits to the data.
The solid lines show the evolution in our hybrid model.
\label{dressler.plot}}
\end{center}}

In our first models, we assume that the star formation rate for the
spiral galaxies is constant in time ($\fstar =0.5$), and that
star formation starts at
$\tstart=0$ for all galaxies.
Figure \ref{es0.plot} shows the evolution of the mean $M/L$ ratio in
this model, for elliptical galaxies (dotted line), S0 galaxies (dashed
line), and elliptical and S0 galaxies combined (solid line). The model
reproduces the observed evolution of the $M/L$ ratio of the combined
sample of elliptical galaxies and S0 galaxies. However, elliptical
galaxies in this model have systematically higher $M/L$ ratios and
redder colors than S0 galaxies at all redshifts.  The difference is
$\approx 0.24$ in $\ln M/L_B$, or $\approx 0.05$ in $U-B$ color. This
difference corresponds to a difference of $\approx 20$\,\% in
luminosity weighted age (e.g., Worthey et al.\ 1994). This large
difference arises because all elliptical galaxies formed their stars
at extremely high redshift in this model, whereas the S0 galaxies are
continuously transformed from star forming galaxies.

This result is in conflict with studies of luminous early-type cluster
galaxies at low and intermediate redshift, which indicate very little
difference in luminosity weighted age between luminous elliptical and S0
galaxies in the cores of rich clusters (e.g., J\o{}rgensen,
Franx, \& Kj\ae{}rgaard 1996, Ellis et al.\ 1997, van Dokkum et al.\ 1998a,
Jones, Smail, \& Couch 2000, Kelson
et al.\ 2000). Other studies have reported
S0 galaxies with younger stellar
populations, but these tend to have low luminosities
(e.g., Caldwell et al.\ 1993, van Dokkum et al.\
1998b, Kuntschner \& Davies 1998).


\vbox{
\begin{center}
\leavevmode
\hbox{%
\epsfxsize=6.5cm
\epsffile{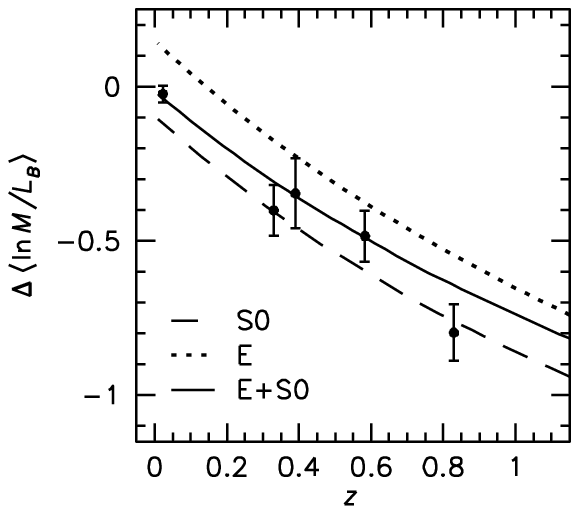}}
\figcaption{\small
Predicted evolution of the mean $M/L_B$ ratio
for a model with different evolutionary histories for elliptical
galaxies and S0 galaxies. It is assumed that all elliptical galaxies
formed at extremely high redshift, whereas S0 galaxies are
continuously transformed from star forming galaxies. This model
predicts a difference of $\Delta \ln(M/L_B) \approx 0.24$ between
luminous elliptical galaxies and S0 galaxies at all redshifts.
\label{es0.plot}}
\end{center}}

There are several ways in which this conflict can be solved.
First, we can adapt the starformation histories.
The  model 
assumes that star formation in elliptical and S0
galaxies commenced at $z=\infty$, and that their star formation
rate was approximately constant from $\tstart$ to $\tstop$.
If star formation in elliptical galaxies commenced
at a later time the systematic difference between elliptical
and S0 galaxies is reduced. If $\tstart = 0.2 \t0$ for
elliptical galaxies (i.e.,
star formation commenced at $z \approx 2.5$)
the systematic difference in $M/L$ ratio
between elliptical and S0 galaxies
at $z=0$ is very small. However,
it increases to $\approx 0.3$ in $\ln (M/L_B)$ at $z=0.8$.
This corresponds to a systematic
difference in  $U-B$ color of $0.06$, which is
inconsistent with the low scatter in the CM relation of
early type galaxies 
at $z=0.83$ (van Dokkum et al.\ 2000).

An alternative explanation for the small observed difference between
luminous elliptical and S0 galaxies
is that most of their
stars were formed at very high redshift.
In our models, this corresponds to low values of $\fstar$,
the parameter describing the star formation history.
The systematic difference in $M/L$ ratio scales linearly with
$\fstar$, and for values of $\fstar \lesssim 0.2$ the
systematic age difference between elliptical and S0 galaxies
is $\lesssim 5$\,\% at all redshifts.
This explanation can be tested with the help of direct measurements
of the star formation rates of the spiral galaxies in intermediate
redshift clusters.

A third possibility is that the effects of metallicity
cancel the effects of age (e.g., Worthey, Trager, \& Faber 1996,
Shioya \& Bekki 1998, Trager et al.\ 2000).
The spiral galaxies would be enriched in metals compared to the 
elliptical galaxies of the same luminosity, and hence their colors 
would be redder, and their $M/L$ ratios higher than that of elliptical
galaxies with the same age.
This solution requires exact fine tuning between age and metallicity
to keep the scatter small. It can be tested by direct observations of
absorption line strengths of elliptical and S0 galaxies in nearby and
distant clusters. The first studies do not appear to indicate
significant differences between massive elliptical and S0 galaxies
(e.g., Kuntschner \& Davies 1998, Jones et al.\ 2000,
Kelson et al.\ 2001).

Finally, the separation of ellipticals and S0's is very difficult,
and it is not clear whether a clear distinction can be made between
elliptical galaxies and S0 galaxies with $L_*$ luminosities
(e.g., J\o{}rgensen \& Franx 1994, Rix \& White 1990). 
The latter study demonstrated that many elliptical galaxies contain
significant disks, and would be classified as S0 galaxies when seen edge-on.
Jorgensen and
Franx argued that the Coma elliptical and S0 galaxies can be approximated
well by a population of galaxies which has a homogeneous distribution
in  the ratio of bulge light to total light. In this case, no clearcut
distinction would exist between S0 galaxies and elliptical galaxies, and
the evolutionary histories would not be so different between the galaxies.
This might also explain the fact that the classifications between different
authors agree well when the total fraction of elliptical and S0 galaxies
is considered, but not when the classes are separated (e.g., Andreon 
1998, Fabricant et al.\ 2000).
Quantitative classifications would be able to resolve this issue, especially
if detailed attention is given to the classification uncertainties
due to projection effects.

\section{Summary and Conclusions}

In this paper we have quantified the effects of morphological
evolution on the observed properties of early-type galaxies in
clusters at $0<z<1$. A manifestation of morphological evolution is the
``progenitor bias'': at high redshift, the progenitors of the youngest
present-day early-type galaxies drop out of the sample. In general,
the observed evolution of the mean luminosity and color of early-type
galaxies is slower than the true evolution of all progenitors of
early-type galaxies, and the scatter in luminosity and color is
smaller.  If morphological evolution is strong the observed mean
evolution is similar to that of a single age stellar population formed
at $z=\infty$, and the scatter is approximately constant with
redshift. These results can be derived analytically (cf.\ Appendix
\ref{equations.sec}).

In traditional models it is difficult to explain the low number
fractions of early-type galaxies in distant clusters and the
homogeneity and slow evolution of their stellar populations
simultaneously. We have shown that the observed constant scatter in
the CM relation and the FP and the slow evolution of the mean $M/L$
ratio are natural consequences of morphological evolution. Observations
of early-type galaxies in clusters at $0<z<1$ are well fitted by our
models, and we can constrain the morphological transformation rate and
the star formation histories of galaxies prior to transformation from
the fits. The observed evolution of the early-type galaxy fraction
constrains the transformation rate, the evolution of the mean $M/L$
ratio constrains the time of onset of star formation, and the scatter
in the color-magnitude relation and the Fundamental Plane constrain
the star formation histories of galaxies prior to transformation.
The best fits are obtained if approximately 50\,\% of
present-day early-type galaxies were transformed from other
morphological types at $z<1$, star formation commenced at
early times ($z>2.5$), and star formation was approximately
constant prior to transformation. Such gradual star formation
histories are consistent with the low numbers of star burst
galaxies seen in intermediate redshift clusters (e.g., Abraham et al.\ 1996,
Balogh et al.\ 1998, Ellingson et al.\ 2000), although some
star bursts could be concealed by dust (Poggianti et al.\ 1999).

The observed color and luminosity
evolution of early-type galaxies can be corrected for the
effects of progenitor bias. We have shown that for a given
morphological transformation rate the progenitor bias is
proportional to the scatter in $M/L$ ratio or color.
The correction is maximal if the
scatter in the CM relation and the FP at $z=0$ is entirely due to age
variations between galaxies.  This maximum correction to the evolution
of rest frame $\Delta \ln \langle M/L_B \rangle$ is approximately
$-0.2 \times z$, and the correction to the evolution of rest frame $\Delta
\langle U-B \rangle$ is approximately $-0.04 \times z$. These corrections
can easily be computed for other rest frame bands using stellar
population synthesis models. As an example, the Worthey (1994)
and Bruzual \& Charlot
(2000) models give $\Delta (U-V) \approx 1.7 \Delta (U-B)$ for
solar metallicity and Salpeter (1955) IMF. Therefore, the correction to
$\Delta \langle U-V \rangle$ is $\sim -0.07 \times z$. The corrections
are upper limits, because variations in dust content and metallicity
probably contribute to the scatter. On the other hand, as pointed
out by, e.g., Trager et al.\ (2000) variations in metallicity and age may
conspire to produce a low scatter. If this is confirmed the corrections
may be larger than the values given here.

Because the mean luminosity and color evolution are underestimated,
the mean age of the stars in early-type galaxies that is derived
from them is overestimated in traditional models. 
We find that the mean luminosity
weighted formation redshift of early-type galaxies can be as low as
$\langle z_* \rangle
= 3.0^{+0.9}_{-0.5}$ for $\Omega_m=0.3$ and $\Omega_{\Lambda}=0$, or
$\langle z_* \rangle = 2.0^{+0.3}_{-0.2}$ for $\Omega_m=0.3$
and $\Omega_{\Lambda}=0.7$, if the scatter in the CM relation
and the FP is entirely caused by age variations.
This result is important, because
previous studies which ignored the effects of progenitor bias
found much earlier formation epochs of the
stars in early-type galaxies (e.g., Ellis et al.\ 1997, Stanford
et al.\ 1998, van Dokkum et al.\ 1998a). The redshift range 2 -- 3
is of interest because it is observationally accessible. In particular,
our results are consistent with the idea that (some) Ly-break
galaxies are the star forming building blocks of present-day
early-type galaxies in clusters (Baugh et al.\ 1998).

In our main analysis we treated the evolution of early-type galaxies
as one class, but we also explored a model which separates the two
classes. In our model, only the S0 galaxies evolve morphologically
from spiral galaxies,  as advocated by
Dressler et al.\ (1997). The model  predicts that colors
and $M/L$ ratios of luminous S0 galaxies in nearby clusters are
systematically offset from those of elliptical galaxies, in the sense
that elliptical galaxies are redder and have higher $M/L$ ratios. This
result is in conflict with studies of bright early-type galaxies in
the cores of rich clusters, which generally find very little
difference between elliptical and S0 galaxies (e.g., Bower et al.\ 1992,
Ellis et al.\ 1997). 
There are several ways to resolve this conflict:
star formation may have started later in the elliptical galaxies, or
most of the star formation in
elliptical and S0 galaxies must have occurred at very early times,
long before morphological transformation to early-type
galaxy. 
Alternatively, there might be a conspiracy between age and
metallicity (e.g., Trager et al.\ 2000) such that S0 galaxies have
systematically higher metallicities than elliptical galaxies of the
same luminosity. This can be tested by absorption line measurements;
current measurements do not indicate such an effect
for cluster galaxies.
Finally, the separation between elliptical and S0 galaxies may not be
very sharp. Rix \& White (1990) showed that many S0 galaxies will be
classified as elliptical galaxies when not viewed edge-on.
J\o{}rgensen and Franx (1994) showed that the elliptical and S0 galaxies
in Coma can be modeled by a population of galaxies with a homogeneous
distribution of bulge-to-total light ratio. Hence, the
galaxies which are classified as elliptical galaxies
may also have evolved since $z=1$.

It is interesting to compare our results to predictions of
semi-analytical galaxy formation models in CDM cosmologies (White \&
Frenk 1991; Cole 1991). 
Because the morphologies of galaxies change quite rapidly
in these models (e.g., Baugh, Cole, \& Frenk
1996, Kauffmann 1996) progenitor bias is an important ingredient.
In particular, the models
provide reasonable fits to
the observed scatter in the CM relation (Kauffmann \& Charlot
1998; Cole et al.\ 2000) and the
evolution of the mean $M/L$ ratio of massive
galaxies (Diaferio et al.\ 2000), even though galaxies in these
simulations have quite complex (star) formation histories. However, in
these complex models it is difficult to isolate and study specific
effects such as the progenitor bias.  The main contribution of our
work is that we use very simple models with only three free
parameters. We show that the effects of progenitor bias can be
constrained directly from the observations, and it is straightforward
to reproduce our results.

The main uncertainty in the present paper is the rate of morphological
evolution. More high resolution, large field studies of distant clusters
are needed to constrain the evolution of the early-type galaxy fraction
better. Detailed simulations are needed to determine the effects
of infall on the observed properties and morphological mix
of cluster galaxies (e.g., Ellingson et al.\ 2000).
Furthermore, studies of the evolution of the
mass function of rich clusters to $z=1$ are necessary to determine
whether high redshift clusters are representative progenitors of
present-day rich clusters. As pointed out by, e.g., Kauffmann
(1995) significant evolution in the mass function
of rich clusters may result in a ``progenitor bias'' at the cluster
level.

\begin{acknowledgements}
We thank the referee, Richard Ellis, for constructive comments
which improved the presentation and clarity of the paper
significantly.
P. G. v. D. acknowledges support by NASA through Hubble Fellowship
grant HF-01126.01-99A
awarded by the Space Telescope Science Institute, which is
operated by the Association of Universities for Research in
Astronomy, Inc., for NASA under contract NAS 5-26555.
\end{acknowledgements}

\begin{appendix}
\section{Analytic Expressions}
\label{equations.sec}
\subsection{Evolution of the Mean Luminosity and Color}
Combining Eq.\ \ref{first.eq} -- \ref{levo.eq} gives the following
approximate expression for the observed
evolution of the mean $M/L_B$ ratio of early-type galaxies:
\begin{equation}
\label{allml.eq}
M/L_B(t) \propto \frac{\taustop \left[ 1 - \exp \left( - \frac{t - 0.1
\t0}{\taustop} \right) \right]}{\int_0^{t-0.1\t0}
\exp \left( - \frac{t'}{\taustop}
\right) \left[ t-(\fstar t'+\tstart)\right]^{-\kappa_B} dt'}.
\end{equation}
Similarly, the color evolution can be approximated by
\begin{equation}
\label{allcol.eq}
\frac{L_B}{L_U}(t) \propto \frac{\taustop \left[ 1 - \exp \left( -
\frac{t - 0.1
\t0}{\taustop} \right) \right]}{\int_0^{t-0.1\t0}
\exp \left( - \frac{t'}{\taustop}
\right) \left[ t-(\fstar t'+\tstart)\right]^{\kappa_B-\kappa_U} dt'}.
\end{equation}
For given $\fstar$ the integrals can be evaluated analytically.

If $\taustop=\infty$ and $\tstart=0$ (i.e.,
in the case of a constant transformation rate) evaluation
of the integral in Eq.\ \ref{allml.eq} gives
\begin{equation}
M/L_B(t) \propto \frac{\fstar(1-\kappa_B)(t-0.1 \t0)}
{t^{1-\kappa_B} - \left[ t- \fstar (t-0.1 \t0) \right]^{1-\kappa_B}}.
\end{equation}
For $t \gg 0.1 \fstar \t0$ this reduces to
\begin{equation}
\label{approx.eq}
M/L_B(t) \propto \frac{\fstar (1-\kappa_B)}{1+(1-\fstar)^{1-\kappa_B}}
t^{\kappa_B}.
\end{equation}
This approximation is accurate to better than
$\sim 3$\,\% for $\fstar \lesssim 0.5$ and
$z \lesssim 1$. As a consequence,
\begin{equation}
\Delta \ln (M/L_B)(t) \approx \kappa_B \ln (t).
\end{equation}
Similarly, the color evolution reduces to
\begin{equation}
\Delta (U-B) \approx (\kappa_U - \kappa_B) \, 2.5 \log (t).
\end{equation}
These expressions are identical to the evolution of a single age
stellar population that was formed at $z=\infty$.

\subsection{Evolution of the Scatter}

For $\tstart=0$
the scatter in $\ln M/L$ at time $t$ can be approximated by
\begin{eqnarray}
\sigma (\ln M/L)(t) &\approx & \kappa \, \sigma \left( \ln (t-\fstar
\tstop)
\right) \nonumber\\
& \approx & \kappa \frac{\sigma (t- \fstar \tstop)}
  {\langle t-\fstar \tstop \rangle}\nonumber \\
& = & \kappa \frac{\fstar \sigma (\tstop)}{t - \fstar
  \langle \tstop \rangle}.
\end{eqnarray}
For a constant morphological transformation
rate the distribution of $\tstop$ is a simple top hat,
bounded by $t_1=0$ and $t_2 = t - 0.1 \t0$. The $1 \sigma$ spread
of a top hat distribution
is $0.29 (t_2-t_1)$, and the mean is $(t_2-t_1)/2$. As a result,
\begin{equation}
\sigma (\ln M/L) = \kappa
 \frac{0.29 \fstar t - 0.029 \t0}
 {(1-0.5 \fstar) t - 0.05 \t0}.
\end{equation}
For $t/\t0 \gg 0.1 \fstar$ (i.e., at low redshift) and
small $\fstar$, this reduces to
\begin{eqnarray}
\label{scatevo.eq}
\sigma (\ln M/L_B) &\approx& \kappa_B \frac{0.29 \fstar}{1-0.5 \fstar} \\
 &\sim & 0.3 \fstar
\end{eqnarray}
in the rest frame $B$ band. Similarly, the scatter in the color
can be approximated by
\begin{eqnarray}
\sigma (U-B)& \approx& 2.5 (\log e) (\kappa_U - \kappa_B)
\frac{0.29 \fstar}{1-0.5 \fstar}\\
& \sim & 0.05 \fstar.
\end{eqnarray}
The scatter is to good approximation independent of redshift, and a
strong function of $\fstar$, the parameter describing
the star formation history of individual galaxies.
\end{appendix}

\end{document}